\documentclass[aps,pre,twocolumn,superscriptaddress,longbibliography]{revtex4-2}

\usepackage[utf8]{inputenc}
\usepackage[T1]{fontenc}
\usepackage{graphicx}
\usepackage{amsmath,amssymb}
\usepackage{bm}
\usepackage{url}
\usepackage{hyperref}

\begin{document}

\title{Crowding controls the scaling of bus frequency with demand}

\author{Siddharth Patwardhan}
\email{siddharthpatwardhan1@gmail.com}
\affiliation{Center for Science of Science and Innovation, Kellogg School of Management, Northwestern University, Evanston, IL 60208, USA}

\author{\c{S}irag Erkol}
\affiliation{Center for Science of Science and Innovation, Kellogg School of Management, Northwestern University, Evanston, IL 60208, USA}

\author{Filippo Radicchi}
\affiliation{Center for Complex Networks and Systems Research, Luddy School of Informatics, Computing, and Engineering, Indiana University, Bloomington, IN 47408, USA}

\author{Marc Barthelemy}
\email{marc.barthelemy@gmail.com}
\affiliation{Universit\'e Paris-Saclay, CEA, CNRS, Institut de Physique Th\'eorique, 91191 Gif-sur-Yvette, France}
\affiliation{Centre d'Analyse et de Math\'ematique Sociales (CNRS/EHESS), 54 Avenue de Raspail, 75006 Paris, France}
\affiliation{Complexity Science Hub, Vienna, Austria}

\begin{abstract}
Cities must allocate limited resources to maintain mobility, with uncertainties about the resulting state of the system. Analyzing roughly 3{,}000 bus routes with more than 4 billion yearly riders across 19 metropolitan areas worldwide, we uncover a robust scaling law of the form $f \sim (d/t)^\alpha$ with exponent $\alpha \in [1/2,\,2/3]$, linking the service frequency $f$ to passenger demand $d$ and route duration $t$.
We show that this scaling emerges from a simple optimization principle: cities implicitly minimize total passenger waiting time under a fixed operational budget when both schedule frequency and crowding are taken into account.
This mechanism produces two universal regimes: a \emph{frequency-dominated} regime with $\alpha = 1/2$ when crowding is negligible, and a \emph{capacity-dominated} regime with $\alpha = 2/3$ when most routes are overloaded.
Intermediate exponents arise when only part of the network operates near capacity. Furthermore, we find that the benefits of additional investment are highly uneven across systems. For instance, our model suggests that a $20\%$ budget increase yields nearly a 5-minute reduction in daily waiting time per passenger in Boston, compared to only about 1 minute in Paris. These findings place urban transit within a broader class of constrained capacity-allocation problems, while highlighting a distinct regime in which prescribed route demands shape the allocation of limited service resources. The resulting scaling laws show how simple optimization principles can generate
systematic exponents in complex transport systems, beyond the dissipation-based frameworks usually considered in physical and biological flow networks.
\end{abstract}


\maketitle

\section*{Significance Statement}
Public transit agencies face the challenge of allocating limited resources across routes with varying demand. Analyzing bus systems across 19 metropolitan areas, we show that, despite differences in cities, agencies, and planning practices, realized operations exhibit a specific scaling pattern in service allocation. This regularity is not imposed by a universal planning formula, but emerges across institutional and urban contexts. We show that it can be understood through a constrained-optimization principle balancing passenger waiting time, crowding, and limited resources. The result connects urban transit to complex flow systems in physics and biology by highlighting a regime where demand fixes flows, and cities allocate service capacity. This framework explains unequal returns to investment across systems and guides efficient, equitable planning.

\section{Introduction}
Urban mobility systems lie at the intersection of efficiency and constraint, a tension that shapes many complex infrastructures~\cite{cats202550,bellocchi2021dynamical}. Every metropolitan area must allocate limited operational resources---vehicles, drivers, energy---to sustain mobility demands that fluctuate in space and time \cite{cats2019frequency,chodrow2016demand}. 
In the case of public transport, and in particular bus networks, allocation of resources is determined by how service frequencies are distributed across routes. 
The question of how cities `decide' these frequencies, consciously or implicitly, touches on a fundamental issue: how collective efficiency emerges from local optimization under resource constraints.

In transport economics, this problem was first formalized through a series of stylized optimization models for a single bus line.
In Mohring’s seminal analysis~\cite{Mohring1972}, the optimal headway arises from balancing operator costs, which grow linearly with service frequency, and users’ waiting costs, which decrease inversely with frequency. 
The resulting condition $f^\star \sim \sqrt{d/t}$---often referred to as the \emph{square-root rule}---reflects a basic scaling relation between demand $d$, trip duration $t$ and optimal service frequency $f^\star$. This square-root form is in fact the only possibility consistent with dimensional analysis when the frequency is given by the number of trips per unit time, demand is expressed in passengers per unit time, $d$, and the route has a given duration $t$. Any other power-law relation, such as $f \sim (d/t)^\alpha$ with $\alpha \neq 1/2$, would imply the existence of an additional temporal or spatial scale in the system, pointing to richer mechanisms or constraints beyond the minimal model.

Subsequent works by Jansson~\cite{Jansson1980} and by Jara-Díaz and Gschwender~\cite{JaraDiazGschwender2003} extended this framework to include vehicle capacity, crowding costs, and multiple lines, showing that similar scaling relations persist even in more realistic settings.
Despite their simplicity, these models remain the theoretical backbone of most frequency-setting approaches, implicitly guiding the design of transit supply in cities worldwide.
At the same time, these models are primarily prescriptive, derived under stylized assumptions.
In practice, bus frequencies are revised through iterative planning processes that account for many factors, such as ridership data, load targets, service standards, accessibility goals, budget constraints, and political considerations \cite{plan_obj1,plan_obj2,plan_obj3,plan_obj4,pol1,pol2}.
The resulting allocations therefore emerge from repeated adjustments across systems with diverse demand profiles, institutional priorities, geographies, and infrastructures.

From a broader perspective, frequency-allocation problems belong to a large family of constrained transport problems: how to allocate limited capacity so as to reduce a global cost, which may be expressed in terms of time, energy, dissipation, or another generalized loss
\cite{banavar1999size,bohn2007structure,durand2007structure,corson2010fluctuations,katifori2010damage}.
In statistical physics and complex network studies, related questions arise in systems ranging from vascular and leaf venation networks to river basins and transportation systems
\cite{laporte2019introduction,aldous2019optimal,patwardhan2024symmetry}.

In many physical and biological examples, the central problem is to understand how flows distribute through a conductance network, and how the conductances adapt under energetic or material constraints. In the standard electrical or hydraulic formulation, edge conductances $g_e$ are chosen under a global
resource constraint, while the edge currents $I_e$ are determined
self-consistently from node sources and sinks through Kirchhoff laws. A common objective is the minimization of dissipated power, which yields to scaling relations between current intensity and conductance where the exponent is controlled by the form of the material constraint
\cite{bohn2007structure,durand2007structure,corson2010fluctuations,katifori2010damage}. Such mechanisms explain why optimized physical and biological networks often
exhibit characteristic relations between flow, conductance, and topology.

Public-transport networks provide an analogous, but distinct, setting. Here, 
passenger demands at the route level, i.e., edge currents, are imposed,
while service frequencies play the role of allocated capacities or effective conductances. The optimization problem is therefore not primarily the redistribution of currents within a conductance network, nor the minimization of physical dissipation, but rather the design of service capacity across routes carrying heterogeneous demand. The leading cost is a user-time cost, dominated in the simplest case by waiting-time-like terms, rather than a dissipated-power
functional (see section 2.2 in the SM for a longer discussion).
This analogy motivates a statistical-physics view of urban transport in which scaling relations between demand, frequency, and performance are interpreted
as signatures of constrained capacity allocation, rather than as mere
engineering heuristics. 

In the present work, we build on the classical Mohring--Jansson framework and reinterpret urban bus networks as complex flow systems that allocate limited service capacity so as to minimize aggregate user cost under resource constraints. 
Within this framework, we investigate how the interplay between frequency, vehicle capacity, crowding, and budget limitations leads to distinct scaling regimes between service frequency and passenger demand.

\section{Results}

\subsection{Empirical results}

\begin{figure*}[t]
    \centering
    \includegraphics[width=\textwidth]{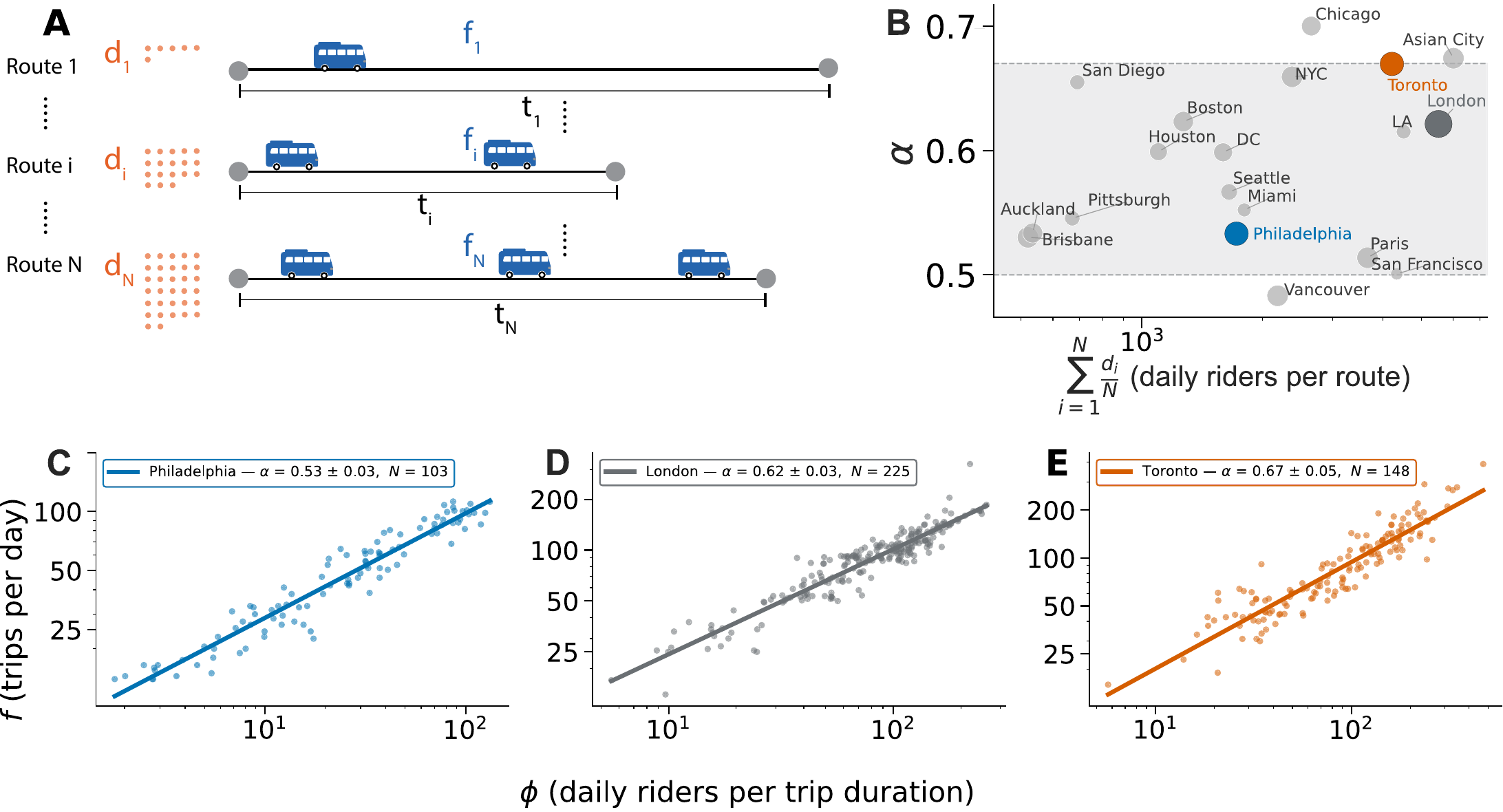}
    \caption{
    \textbf{Scaling of resource allocation in bus systems across the world.}
    (A) Schematic representation of the route-level quantities.
    Each route $i$ is characterized by the daily passenger demand $d_i$, the daily number of trips or service frequency $f_i$, and the route duration $t_i$. Orange dots depict passengers, blue bus icons show service frequency, and the horizontal bracket denotes the time required to complete the route.
    The scaling exponent $\alpha$ quantifies how $f$ grows with $\phi = d/t$ according to $f \sim \phi^{\alpha}$. 
    (B) City-specific scaling 
    exponents $\alpha$ for $19$ urban bus systems plotted against the average number of daily riders per route. 
    The shaded band denotes the 
    range $1/2 \le \alpha \le 2/3$ predicted by our theoretical model.
    (C-E) Representative 
    city-specific scaling relationships.
    Each point represents one of the $N$ single bus routes serving the corresponding city. Solid lines show the best-fit for the scaling $f \sim \phi^{\alpha}$, obtained by performing simple linear regression for $\log f$ {\it vs.} $\log \phi$. We measure: (C) $\alpha = 0.53\pm0.03$ and $R=0.96$ for Philadelphia, (D) $\alpha = 0.62\pm0.03$  and $R=0.94$ for London, and (E) $\alpha = 0.67\pm0.05$ and $R=0.92$ for Toronto.
}
    \label{fig:1}
\end{figure*}

We compile data for more than 3,000 bus routes from 19 cities, spanning a wide range of sizes, geographies, and operating practices.
A schematic description of the type of data we analyze is provided in Fig.~\ref{fig:1}A; 
a more detailed description of the data sources and preprocessing is provided in the Methods section and the SM.
We first examine how route frequency $f$ scales with route `busyness' $\phi = d/t$ across the cities, where $d$ is the daily ridership and $t$ is the trip duration. 
For each system, we estimate the exponent $\alpha$ from the relation $f \sim \phi^{\alpha}$.
Across all cities, we find that the scaling exponents lie consistently between $\alpha = 0.50$ and $\alpha = 0.67$ except for two outliers (see Fig. \ref{fig:1} B).
The scaling law is robust, holding over almost three orders of magnitude in $\phi$, from lightly used suburban lines to the most crowded urban corridors.
The cities analyzed in this study differ widely in size, geography, and operating practices, yet their exponents fall within a remarkably narrow range.
This points to a common underlying principle in how transit agencies allocate service frequency in response to passenger demand. 
Moreover, the differences between cities are not random. 
Notice that larger or denser systems (e.g., Chicago, New York City) tend to exhibit higher $\alpha$, while smaller or less congested networks (e.g., Vancouver, Philadelphia) remain closer to the square-root regime.
Furthermore, the exponent $\alpha$ is also positively correlated with the total ridership and transit agency size (refer to SM for more details). Furthermore, these exponents appear to be temporally robust, as seen by our analysis of the ridership and scheduling data for the past ten years in Boston and Chicago (see SM).
In both cases, we find that the scaling law persists despite dramatic changes and disruptions in the overall ridership and operations, confirming that the frequency-demand relation is a structural feature of system-level resource allocation.  
The results are robust to alternative regression methods and moderate changes in filtering thresholds. 
The regularity in the range of observed exponents, $1/2 \le \alpha \le 2/3$, motivates a theoretical model explaining this range as a natural outcome of optimizing service under fixed resource constraints.

\subsection{Theoretical model: crowding effect}

The consistent scaling $f \sim \phi^{\alpha}$ observed across cities motivates the question: why do bus systems, despite vast differences in size, geography, and operation, 
display scaling exponents between $1/2$ and $2/3$? 
A positive correlation between the demand per route and $\alpha$ suggests a universal mechanism balancing efficiency and overcrowding (see Fig. \ref{fig:1}). 
To explore this trade-off, we model a transit system that minimizes the total waiting time of passengers under a fixed budget. 

In this model, each route direction $i$ is treated separately and characterized by its daily demand $d_i$, trip duration $t_i$, and service frequency $f_i$.
The total waiting time on a route combines two components: an average waiting time due to the service frequency, 
given by $\frac{1}{2f_i}$ accounting for random arrivals, and an additional crowding penalty that increases with the passenger load per vehicle. 
When the number of passengers approaches the bus capacity, some riders must wait for the next vehicle, effectively increasing the average waiting time. To model this effect, we denote by $\tau$ the average time a passenger spends onboard, and by $\kappa$ the vehicle capacity.

On average, each passenger spends a time $\tau$ riding on route $i$ over an operating period of length $t_i$, so the probability that a given passenger is on the bus at an arbitrary time is $\tau/t_i$. This implies that the average number of passengers simultaneously present on route $i$ is $d_i\,\tau/t_i$. 
The number of trips per day on the line is $f_i$, thus
the average load per bus is 
\[
\langle n_i\rangle = \frac{d_i\,\tau}{t_i\,f_i}\,.
\]
We compare this quantity to the vehicle capacity $\kappa$ and assume that crowding generates an additional waiting-time contribution proportional to the load–capacity ratio, of order $\langle n_i\rangle/\kappa$. The magnitude of this term determines whether we are in a frequency-limited regime 
 ($\langle n_i\rangle < \kappa$)
or in a capacity-limited regime ($\langle n_i\rangle > \kappa$).
This argument is in the spirit of congestion effects on roads (see, e.g., \cite{branston1976link}). 
We thus model the total waiting time on route $i$ (cf.\ similar expressions in \cite{Jansson1980}) as
\begin{equation}
    C_i(f_i) 
    = \frac{d_i}{{2}f_i}\left[\,1 + \delta\,\frac{d_i}{f_i t_i}\right],
\label{eq:cost}
\end{equation}
where $\delta=\tau/\kappa$ is 
the parameter that tunes the strength of crowding effects. The cost function of Eq.~(\ref{eq:cost}) measures the average waiting time per day, i.e., a dimensionless quantity. This expression makes explicit how waiting time first decreases with increasing frequency and then rises more sharply when passenger loads approach vehicle capacity.

The operational budget is defined as the total `driving time' per day in the system,
\begin{align}
    B = \sum_{i=1}^{N} f_i\, t_i ,
\end{align}
where $N$ is the number of routes. 
This quantity represents the total number of daily vehicle-hours operated 
, with $f_i$ capturing the number of daily vehicle trips supplied on route $i$ and $t_i$ capturing the operating time associated with each trip, including driver time and fuel use.
It therefore provides a proxy for the total cost associated with staffing, fleet deployment, and fuel use.
The optimal solution that minimizes the total waiting time $\sum_i C_i(f_i)$ subject to this constraint yields the optimality condition
\begin{equation}
\phi_i f_i^{\star-2} + 2\delta\,\phi_i^2 f_i^{\star-3} = \lambda,
\label{eq:phi}
\end{equation}
where $\phi_i = d_i/t_i$ and $\lambda$ is the Lagrange multiplier enforcing the budget constraint (substituting the solution into the budget equation then determines $\lambda$ as a function of $B$, $\{d_i\}$ and $\{t_i\}$). 
The expression in Eq.~(\ref{eq:phi}) reveals two asymptotic regimes for the optimal frequency $f^\star_i(\phi_i)$.

When crowding is negligible, the first term dominates, and the model reduces to the classical square-root law 
$f_i^* \sim \phi_i^{1/2}$. The exponent $1/2$ arises here from minimizing a waiting-time-like cost under a fixed service budget. This mechanism should not be confused with the scaling exponents that appear in dissipative transport networks, where the objective is instead to minimize power dissipation under material or capacity
constraints~\cite{bohn2007structure,durand2007structure,corson2010fluctuations}. In such systems, the relation between flow and conductance depends on the physical constraint imposed on the network, rather than on a waiting-time objective. A related distinction is emphasized by Jensen et al.~\cite{jensen2013optimal}, who showed that optimal transport in concentration-limited systems results from a balance between increased transported material and increased impedance, with the optimum depending on the imposed physical constraint, such as constant pressure or constant work rate. The common point is therefore not the microscopic cost being minimized, but the broader structure of a constrained allocation problem in which limited
capacity is distributed across heterogeneous flows or demands.

When crowding effects dominate, the second term prevails, and we obtain 
$f_i^* \sim \phi_i^{2/3}$.
The crossover between these two regimes accounts for the range of exponents observed in real systems, see Fig.~\ref{fig:1} B. The analysis of Eq.~\ref{eq:phi} shows that the relevant scaling variable is 
$y = 2\sqrt{\lambda\,\delta\,\phi}$, indicating that the two regimes are connected by a smooth crossover (see SM). This crossover is characterized by a threshold value $\phi_c$ given by
\begin{equation}
    \phi_c = \frac{1}{4\lambda\,\delta^{2}}.
\end{equation}
Routes with $\phi_i < \phi_c$ operate in the frequency-dominated regime, while those with $\phi_i > \phi_c$ lie in the capacity-dominated regime. 
Because $\phi_c$ depends on both the total resources available (through $\lambda$) and the severity of crowding (through $\delta$), different cities 
are characterized by different crossover values.
When a system spans both sides of the crossover, the behavior can be described by an effective exponent $\alpha$ that falls 
anywhere between $1/2$ and $2/3$.

\begin{figure*}[t]
    \centering
    \includegraphics[width=\textwidth]{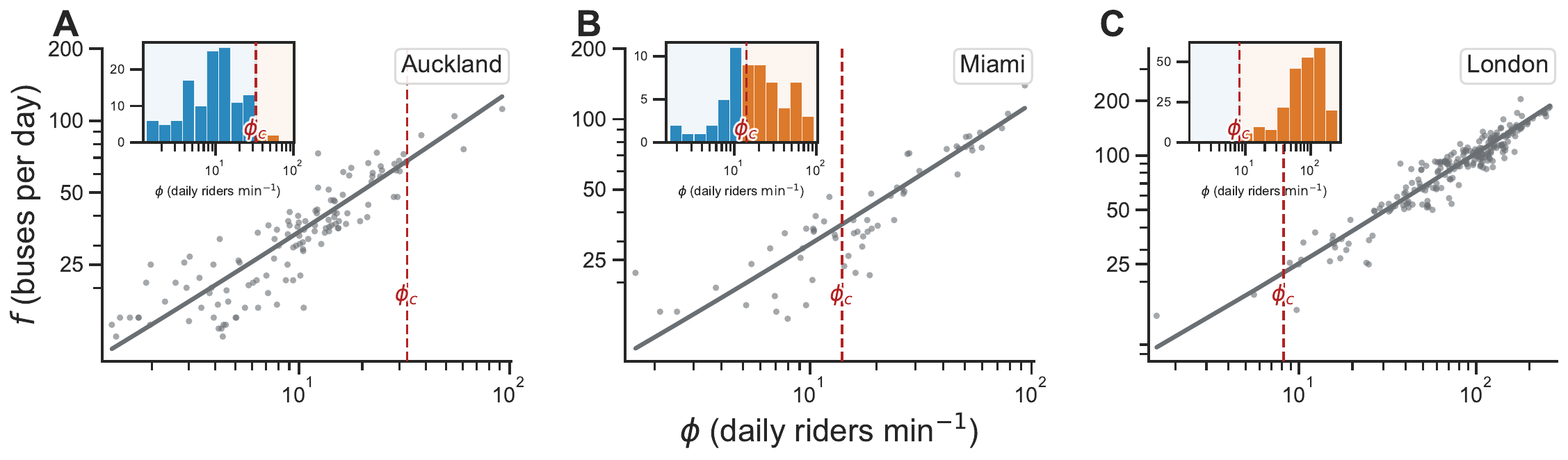}
    \caption{
    \textbf{Frequency-dominated {\it vs.} capacity-dominated routes}.
(A) We plot the frequency of service $f$ as a function of the ratio $\phi = d/t$ for each bus route serving the city of Auckland.
The gray curve corresponds to the best fit of our model with the empirical data points. Best estimates of the model parameters are $\delta=0.78$ minutes of bus operation per passenger and  $\phi_c=32.7$ daily riders per minute.
The vertical dashed line marks the theoretical crossover scale $\phi_c$, separating the frequency-dominated regime ($\phi < \phi_c$) from the capacity-dominated regime ($\phi > \phi_c$). 
Inset: distribution of $\phi$ values highlighting the relative share of routes in each regime. Almost all routes are in the frequency-dominated regime; this explains why the effective exponent of the scaling relation $f \sim \phi^\alpha$ is  $\alpha=0.53\pm 0.05$.
(B) Same as in (A) but for Miami, with $\delta=0.95$ and crossover threshold $\phi_c=14.0$. Miami is characterized by a similar number of routes operating below and above the threshold, resulting in an effective scaling exponent $\alpha=0.55\pm0.05$.
(C) Same as in (A) but for London, with 
$\delta=1.04$
and crossover threshold $\phi_c=8.2$.
Most of the routes in London are in the capacity-dominated regime, and the effective scaling exponent is $0.62\pm0.03$.
}
    \label{fig:2}
\end{figure*}

To compare the model with real systems, we infer the crowding parameter $\delta$ by fitting the model-predicted frequencies to the observed route frequencies for each city.
This fitted value of $\delta$, together with the empirical budget constraint, determines the crossover threshold $\phi_c$ and therefore the operating regime of each route.
Fig.~\ref{fig:2} illustrates this crossover in real systems, showing three representative cities.
In Fig.~\ref{fig:2}A, the solid line displays the best fit of the proposed model for Auckland, with the vertical dashed line marking the crossover threshold $\phi_c$. In this case, most routes lie below $\phi_c$, operating in the frequency-dominated regime.
In Fig.~\ref{fig:2}B, we show the best fit for Miami, where the crossover clearly separates routes below $\phi_c$ from those above it, the latter operating in the capacity-dominated regime.
Finally, Fig.~\ref{fig:2}C shows the results for London. London’s routes lie predominantly above $\phi_c$, indicating a largely crowding-dominated system.
The exponent of the power-law fit increases from values near $1/2$ to values closer to $2/3$ (here $\sim 0.62$ for London), in agreement with our theoretical analysis.
We remark that our analysis provides an aggregate view of capacity pressure, rather than assuming that crowding is uniform throughout the day.
Using the fine-grained passenger data from the anonymized Asian city \cite{asian_city_ridership}, we find that hourly service frequency and demand vary substantially over the course of the day.
Moreover, across routes, maximum hourly demand and frequency display a strong linear correlation to their corresponding average values.
Thus, the average quantities used in our model seem to preserve the broad ordering of routes by peak demand and peak service.
We report these results in the SM, along with an empirical validation of the inferred operating regime for the Asian city.

Our results build on and complement a long line of work in transport economics. 
In Mohring’s classical analysis, optimizing frequency under operator costs and purely waiting-time costs yields the square-root rule \cite{Mohring1972}. 
Subsequent models, notably Jansson’s joint optimization of bus size and frequency on a single line \cite{Jansson1980} and the extensions by Jara-Díaz and Gschwender \cite{JaraDiazGschwender2003}, broaden this framework by allowing vehicle size to adjust and by explicitly including crowding disutility in the generalized passenger cost. 
These studies suggest that once capacity and crowding are taken into account, the elasticity of the optimal frequency with respect to demand need not be fixed at $1/2$; in these models, it can lie anywhere in the interval $[1/2,1]$. 
Our contribution is twofold. 
Through a large-scale data analysis effort, we show that real systems concentrate in a much narrower band, $1/2 \leq \alpha \leq 2/3$, when observed at the scale of entire systems. 
Theoretically, our crossover formulation identifies a specific crowding term and global constraint that naturally produce this narrow range by forcing the effective exponent to interpolate between a waiting-dominated value $1/2$ and a crowding-dominated value $2/3$.
Note that a relatively simple modification of our model can also increase the range of allowed exponents to $[1/2, (\beta+1)/(\beta+2)]$, with the tunable parameter $\beta$ that characterizes the nonlinear aversion to 
crowding
The results presented here are valid for $\beta = 1$ (see SM for more details). 

\subsection{Shifting routes toward the frequency-dominated regime}

Within our framework, the two regimes have a clear operational interpretation. 
Routes with $\phi < \phi_c$ lie in the frequency-dominated regime, where the average waiting time is governed primarily by the service frequency. 
Conversely, routes with $\phi > \phi_c$ operate in the capacity-limited regime, where most waiting time arises from vehicle crowding and insufficient capacity. 
From the standpoint of passenger comfort, reliability, and resilience to demand fluctuations, the frequency-dominated regime is generally preferable, as buses are not systematically overloaded and additional demand can be absorbed without excessive crowding.

The model proposed here also clarifies how agencies can shift routes from the capacity-dominated regime into the frequency-dominated one. 
In our formulation, the crossover threshold $\phi_c$ depends on two key factors: the total operational budget $B$ available for service and the effective vehicle capacity (measured here by the parameter $\delta=\tau/\kappa$). 
Increasing either the service budget or vehicle capacity raises $\phi_c$, enabling more routes to operate in the frequency-dominated regime and thereby reducing aggregate passenger costs for a given demand pattern. 
To quantify these effects in real systems, we conduct an experiment in which we hypothetically increase each city's budget (or capacity), while keeping route lengths and demands constant. 
For each scenario, we recompute the optimal frequencies using the best-fit value for $\delta$ found from our model and determine whether individual routes belong to either the frequency- or the capacity-dominated regimes.
Note that the additional budget is not distributed uniformly across routes.
For each budget level, we solve the global optimization problem using the inferred value of $\delta$ and recompute the optimal frequencies for every route.
The resulting allocation naturally assigns disproportionately larger frequency increases to busier routes, as described in more detail in the SM.

We track the evolution of the share of routes in the frequency-dominated regime as a function of the percentage increase in the budget, and this analysis reveals three broad categories of cities. 
First, frequency-dominated cities where most of the routes operate in the frequency-dominated regime, thus not requiring further improvement. Examples of such systems include smaller cities like Philadelphia, Auckland, and Pittsburgh. 
Second, capacity-dominated cities where a large fraction of routes are in the capacity-limited regime. Those cities would require a substantial increase of resources before a significant fraction of their routes could escape the capacity-dominated regime. Cities such as New York, London, and Chicago fall into this second category.
Between these extremes, a group of intermediate cities sits close to the crossover point and is therefore highly sensitive to modest changes in resources. This group comprises cities like Miami, Houston, and Seattle. 
In these cities, relatively small budget increases can shift a large fraction of routes out of the capacity-dominated regime. 

\begin{figure}[t]
    \centering
    \includegraphics[width=\linewidth]{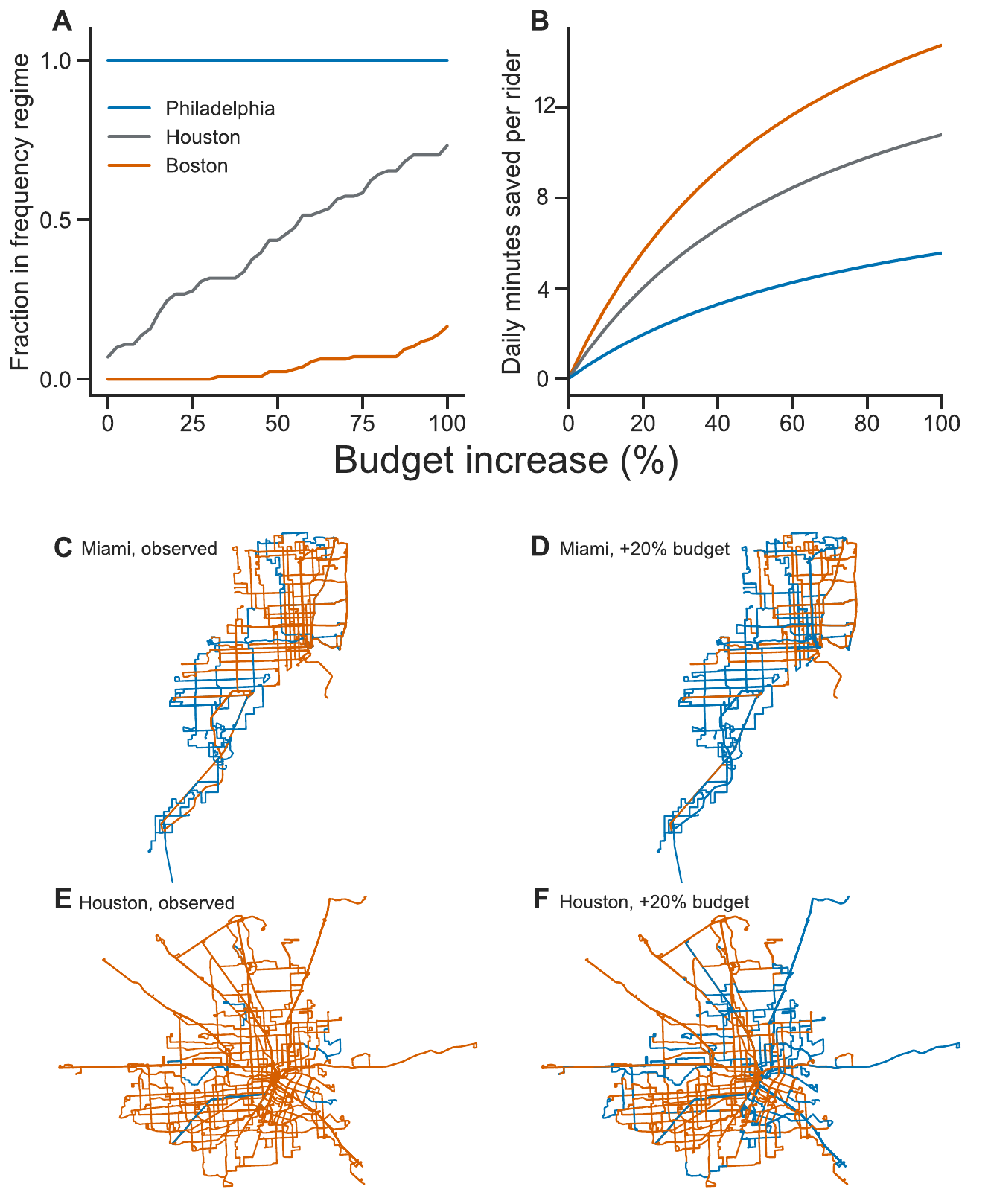}
    \caption{
    \textbf{Potential for improvement in urban transit systems.}
(A) Fraction of routes in the frequency-dominated regime as a function of the budget increase for three representative systems: Philadelphia (blue), Boston (orange), and Houston (gray).
(B) Minutes saved per passenger per day as a function of the budget increase for the three representative systems.
An 18-hour operation window is assumed for all routes to calculate the time saved.
The marginal gain here (slope at the origin) for Philadelphia, Houston, and Boston is $0.11$, $0.24$, and $0.34$,  daily minutes saved per rider per \% of budget increase, respectively.
(C-F) Spatial distribution of route regimes for Miami (C and D) and Houston (E and F). 
Blue and orange lines denote routes in the frequency-dominated and capacity-dominated regimes, respectively. 
Panels C and E show the observed systems, whereas panels D and F show the model-optimized systems obtained by assuming a 20\% budget increase.}
\label{fig:3}
\end{figure}

Figure~\ref{fig:3} summarizes these results. 
In panel~\ref{fig:3}A, we show the fraction of routes in the frequency-dominated regime for three representative cities as a function of budget increase. 
Frequency-dominated cities, such as Philadelphia, exhibit curves that are nearly flat and close to 1, reflecting their already favorable operating conditions. 
Capacity-dominated cities, like Boston, start at low baseline values and increase only slowly with additional resources. 
Intermediate cities, such as Miami, exhibit steep initial slopes, indicating strong responsiveness to modest investment. 
Conversely, cities in the capacity-dominated regime exhibit substantially larger marginal reductions in generalized passenger time, owing to the stronger crowding penalty.
Figure \ref{fig:3}B reports the daily minutes saved per passenger for the three cities as a function of budget increase, computed over an 18-hour operating period.
Figure~\ref{fig:3}C and E show the baseline spatial distribution of regimes for two cities in the intermediate regime, Miami and Houston, respectively. 
Panels~\ref{fig:3}D and F track how the geography of routes in the capacity regime evolves under a budget increase of $20\%$ for both cities (more details in SM). 
Relative capacity increase shows qualitatively similar results (see SM). 

\section{Discussion}

In this work, we compiled and analyzed a large, heterogeneous dataset of bus operations and ridership, covering thousands of routes and billions of passenger trips across $19$ cities worldwide. 
This effort revealed a strikingly tight scaling relation $f \sim \phi^\alpha$ between frequency of service $f$ and the ratio of demand and route duration $\phi = d/t$, with exponents concentrated in the range $1/2 \le \alpha \le 2/3$. 
This regularity is especially surprising because service planning varies widely across agencies in methods, constraints, and priorities, while combining many competing factors such as ridership, load standards, reliability, accessibility, equity, budgets, public input, and political pressures~\citep{plan_obj1,plan_obj2,plan_obj3,plan_obj4,pol1,pol2}.
We then proposed a simple optimization model that explains this scaling as the result of a crossover between a frequency- and a capacity-dominated regime. The theoretical model also provides a practical tool for exploring how changes in budget or capacity would redistribute service. 
Together, these elements connect a new empirical regularity to an underlying mechanism and show how it can inform resource-allocation decisions in urban public transport. 
Such improvements in service quality may also have broader system-level effects, for example, by shifting demand toward public transport and easing pressure on urban road networks.

More broadly, the emergence of these exponents suggests an analogy with physical and biological transport networks, while also highlighting an important distinction. In many optimal electrical, hydraulic, or biological networks, flows and conductances are coupled self-consistently and the optimization is often based on dissipation under material constraints. Urban transit instead starts from largely imposed route-level passenger demand and allocates limited service capacity across routes. Bus systems therefore provide a concrete example of how constrained resources and heterogeneous
loads can generate systematic scaling laws in a capacity-design problem.

\section{Materials and Methods}

\subsection{Data}

For this study, we compiled a unique dataset based on two types of data: (a) route-level ridership data from $19$ cities worldwide and (b) their corresponding General Transit Feed Specification (GTFS), a standardized format that describes public transit schedules and routes.

The most recent monthly, yearly, or seasonal ridership reports published by the relevant transit agencies are used to obtain the daily route demand \cite{chicago_ridership_cta, boston_ridership, philapelphia_ridership, pittsburgh_ridership, losangeles_ridership, sanfrancisco_ridership, houston_ridership, asian_city_ridership,seattle_ridership,london_ridership,dc_ridership,auckland_ridership,brisbane_ridership,miami_ridership,nyc_ridership,toronto_ridership,vancouver_ridership,paris_ridership,sandiego_ridership}
We represent each route by its service in a single direction. 
Because published ridership data are typically reported at the route level and do not distinguish between directions, we infer one-direction demand by assuming that daily ridership is approximately split evenly between the two directions of a route.
For each route, we use the average weekday ridership to estimate its total daily demand, denoted by the number of daily passengers $d$. 
Further details, including the data-collection period, the number of routes per city, and the distributions of ridership, route length, and frequency, are provided in the SM.

We rely on the Transitland data platform \cite{transitland_gtfs} for the GTFS files from the transit agencies. 
The most accurate archived version of the data is used to match the ridership data used in the analysis. 
The GTFS data is used to infer the daily weekday frequency ($f$) and the total time required to complete a single trip ($t$) for each route along a single direction. 
Again, we assume that the frequencies and route durations are symmetric between the two directions of travel to infer $f$ and $t$ along a single direction.
We calculate the total trip duration for each route using the data from the first non-holiday Tuesday of the month. 
This method ensures consistency across all cities. 
Finally, the median of all its trip durations on the selected day is used as the route's travel time, $t$. 
We only consider routes with $d > 25$ passengers per day, $t > 10$ minutes, and $f > 12$ buses per day to filter out infrequently used or erroneous routes.
The analysis of the datasets is restricted to conventional bus services.
All non-traditional routes, for instance, streetcar routes in Toronto and San Francisco, and school buses in Brisbane, London, and Auckland, are excluded from the analysis due to their distinct operational nature. 

\section*{Author Contributions}
S.P. performed the analyses. All authors contributed to the conception and design of the study, interpretation of the results, and writing of the paper. \\\\

\section*{Competing Interests}
The authors declare no competing interests.

\section*{Data and Code Availability}
Processed data and the code developed are available at \url{https://github.com/SiddharthP96/urban-bus-scaling-law}.

\begin{acknowledgments}
This work received partial support from the Air Force Office of Scientific Research (Grant No. FA9550-24-1-003).
The funders had no role in study design, data collection, and analysis, the decision to publish, or any opinions, findings, conclusions, or recommendations expressed in the manuscript.
\end{acknowledgments}

\bibliography{bibliography}

\end{document}